\documentstyle[preprint,aps]{revtex}
\begin{document}
\draft
\title{Transition from localized to extended eigenstates
in the ensemble of power-law random banded matrices
}
\author{  Alexander D. Mirlin$^1\dagger$, Yan V. Fyodorov$^2\dagger$,\\
Frank-Michael Dittes$^3$,  Javier Quezada$^4$ and Thomas H. Seligman$^5$
}
\address{
$^1$ Institut f\"{u}r Theorie der Kondensierten Materie,
  Universit\"{a}t Karlsruhe, 76128 Karlsruhe, Germany}
\address{ $^2$ Fachbereich Physik, Universit\"at-GH Essen, 45117 Essen,
Germany}
\address{ $^3$ Forschungszentrum Rossendorf, Institut f\"ur
Kern- und Hadronenphysik, 01314 Dresden, Germany}
\address{ $^4$ Tecnologico de Monterey, Guadalajara Campus,
Guadalajara, Mexico}
\address{ $^5$ IFUNAM Laboratorio de Cuernavaca, 62191 Cuernavaca,
Morelos, Mexico}
\date{\today}
\maketitle
\tighten
\begin{abstract}
We study statistical properties of the ensemble of
large $N\times N$ random matrices whose entries $ H_{ij}$
decrease in a power-law fashion $H_{ij}\sim|i-j|^{-\alpha}$.
Mapping the problem onto a nonlinear $\sigma-$model with non-local
 interaction, we find a  transition from localized to
 extended states  at $\alpha=1$. At this critical value of $\alpha$
the system exhibits multifractality and spectral statistics
intermediate between the Wigner-Dyson and Poisson one. These features
are reminiscent of those typical for the mobility edge of disordered
 conductors. We find a continuous set of critical theories at $\alpha=1$,
parametrized by the value of the coupling constant of the $\sigma-$model.
At $\alpha>1$ all states are expected to be localized
 with integrable power-law tails. At the same time, for $1<\alpha<3/2$
the wave packet spreading at short time scale is superdiffusive:
$\langle |r|\rangle\sim t^{\frac{1}{2\alpha-1}}$, which leads
 to a modification of the Altshuler-Shklovskii behavior of the
 spectral correlation function.
At $1/2<\alpha<1$ the statistical properties of eigenstates are similar
to those in a metallic sample
in $d=(\alpha-1/2)^{-1}$ dimensions. Finally, the region $\alpha<1/2$
is equivalent to the corresponding Gaussian ensemble of
 random matrices $(\alpha=0)$.
The theoretical predictions are compared with
 results of numerical simulations.
\end{abstract}
\pacs{PACS numbers: 71.30.+h, 71.55.Jv, 05.45.+b, 72.15.Rn}
\narrowtext

\section{Introduction.}
Recently, there has been a considerable interest in the
 properties of large $N\times N$ random banded matrices (RBM).
The ensemble of RBM is defined as the set of matrices with elements
\begin{equation}\label{1a}
H_{ij}=G_{ij}a(|i-j|)\ ,
\end{equation}
 where the matrix $G$ runs over the Gaussian Orthogonal Ensemble (GOE),
and $a(r)$ is some function satisfying the condition $\lim_{r\to
\infty}a(r)=0$
and determining the shape of the band. In the most frequently
considered case
of RBM the function $a(r)$ is considered to be fast (at least,
exponentially)
decaying when $r$ exceeds some typical value $b$ called the bandwidth.
Matrices of this sort were first introduced as an
 attempt to describe an intermediate level statistics
 for Hamiltonian systems in a transitional regime between complete
integrability and fully developed chaos \cite{sel} and then appeared
in various contexts ranging from atomic physics (see \cite{flam}
and references therein) to solid state physics \cite{RBMrev} and
especially
in the course of investigations of the quantum behavior of
periodically driven
Hamiltonian systems \cite{is1}. The mostly studied system of the
latter type
is the so-called quantum kicked rotator (KR) \cite{is2} characterized
 by the Hamiltonian
\begin{equation}\label{2a}
\hat{H}=\frac{\hat{l}^2}{2I}+
V(\theta)\sum_{m=-\infty}^{\infty}\delta(t-mT)\;,
\end{equation}
where $\hat{l}=-i\hbar\,\partial/\partial\theta$
is the angular momentum operator
conjugated to the angle $\theta$. The constants $T$ and $I$
are the period of kicks and the moment of inertia, correspondingly,
and $V(\theta)$ is usually taken to be $V(\theta)=k\cos{\theta}$.
Classically, the KR exhibits an unbound diffusion in the angular momentum
space when the strength of kicks $k$ exceeds some critical value.
It was observed, however, that in a quasi-classical regime
 quantum effects suppress the classical diffusion \cite{is2}
 in close analogy with the effect of Anderson localization of a
 quantum particle by a random potential \cite{fish}.

It is natural to consider
the evolution (Floquet) operator $\hat{U}$ that relates values
 of the wavefunction over one period of perturbation,
$\psi(\theta,t+T)=\hat{U}\psi(\theta,t)$,
in the ``unperturbed'' basis of eigenfunctions of the operator
$\hat{l}$: $|l\rangle=\frac{1}{(2\pi)^{1/2}}\exp(i n \theta)$, $n=\pm 0,
\pm 1,\ldots$. The matrix elements $\langle m|U|n\rangle$ tend to
zero when
$|m-n|\to \infty$. In the case $V(\theta)=k\cos{\theta}$ this decay
is faster
than exponential when $|m-n|$ exceeds $b\approx k/\hbar$, whereas
within the band of the effective width $b$ matrix
 elements prove to be pseudorandom \cite{is2}. All these observations
attracted much interest to the statistical properties of the ensemble of RBM
in order to use the extracted information for understanding the dynamical
properties of the KR.

Let us note, however, that the fast decay of $\langle m|U|n\rangle$
in the above mentioned situation is due to the infinite
 differentiability of $V(\theta)=\cos{\theta}$. If we took a function
$V(\theta)$ having a discontinuity in a derivative of some order the
corresponding matrix elements of the evolution operator
 would decay in a power-law fashion when $|n-m|\to
\infty$\footnote{The authors are grateful to F. Izrailev for
 attracting their attention to this fact.}.
In fact, there is an interesting example of a
 periodically driven system where the matrix
elements of the evolution operator decay in a power-law way,
 namely the so-called Fermi accelerator \cite{jose}.
This system does not show the effect of dynamical localization
in  energy space typical for the KR.
 One can hope to understand this difference in behavior studying
properties
of RBM with power-law decay of the function $a(r)$ at infinity.

One may also consider the random matrix (1) as  the Hamiltonian
of a one-dimensional tight-binding model with long-ranged off-diagonal
disorder (random hopping). A closely related problem with non-random
long-range hopping and diagonal disorder was studied numerically
in Ref. \cite{oono}. The qualitative effect of weak long-range
 hopping on the localized states in a
$3D$ Anderson insulator was discussed by Levitov \cite{levitov}.
We will discuss
the correspondence between our results and those of Ref.
\cite{levitov}
later on. Let us note that similar models with a power-law hopping appear
also in other physical contexts \cite{misc}.

As it was shown in Ref. \cite{FM,RBMrev} the conventional RBM model can be
mapped
onto a $1D$ supermatrix non-linear $\sigma-$model, which allows for
 an exact analytical solution. The same $\sigma-$model was derived
initially
for a particle moving in a quasi-$1D$ system (a wire) and being subject to
a random potential. All states are found to be asymptotically localized,
with the localization length
\begin{equation}\label{1}
\xi=\frac{B_2(2B_0-E^2)}{8B_0^2}\propto b^2;
\quad B_k=\sum_{r=-\infty}^{\infty}
a^2{(r)}r^k    \;.
\end{equation}
In the present paper we consider the case of a power-like shape of
the band:
\begin{equation}\label{2}
a(r)\propto r^{-\alpha}\quad \mbox{for large}\  r \;.
\end{equation}
Under these conditions, the derivation of the $\sigma-$model
presented in \cite{RBMrev,FM} loses its validity.

Reconsidering this derivation, we arrive at a more general $1D$
$\sigma-$model with long-range interaction. Performing its perturbative
and renormalization group analysis, we find that the model is much
richer
than the conventional short-range one. In particular,
 it exhibits an Anderson localization transition at $\alpha=1$.
The main scope of the present work is to study the
statistical properties of the model in the whole range of the parameter
$\alpha$.

\section{The power-law random banded matrix ensemble and the
effective non-linear $\sigma-$model}

Let us consider the ensemble given by eq. ({\ref{1a}) with the
function $a(r)$
having the form
\begin{equation}
a(r)=\left\{\begin{array}{c} 1,\quad r\leq b\\(r/b)^{-\alpha},\quad r>b
\end{array}\right.  \;.
\label{3}
\end{equation}
The parameter $b$ will serve to label the critical models with
$\alpha=1$.
We will consider $b$ to be large: $b\gg 1$, in order to justify formally
the derivation of the $\sigma-$model. We will argue later on that
 our conclusions are qualitatively valid for arbitrary $b$ as well.
We will call the ensemble (\ref{1a}), (\ref{3}) the power-law random
banded matrix (PRBM) model.

In Ref.  \cite{FM,RBMrev} it was shown that the RBM model
 (\ref{1a})
can be mapped, for arbitrary bandshape $a(r)$, onto a
field-theoretical model
of interacting $8\times8$ supermatrices $\sigma_{i}\,,\, (i=1,2,...,N)$
characterized by the action
\begin{equation}\label{4}
{\cal S}\{\sigma\}=\mbox{Str}\left\{\frac{1}{2}\sum_{i,j}R_{ij}\sigma_{i}
\sigma_{j}+\sum_{i}U(\sigma_i)\right\} \;.
\end{equation}
Here $R_{ij}=(A^{-1})_{ij}-\frac{1}{N}\delta_{ij}\sum_{kl}(A^{-1})_{kl}$;
$A$ is the matrix with elements $A_{ij}=a^2(|i-j|)$,
\begin{equation}\label{5}
U(\sigma)=\frac{1}{2}\mbox{Str}
\left\{\ln{(E-\sigma-i\frac{\omega}{2}\Lambda)}
+\frac{1}{N}\sum_{k,l}(A^{-1})_{kl}\sigma^2\right\}\ ,
\end{equation}
$\omega$ is the frequency, and $\Lambda={\rm diag}(I,-I)$ in the
 ``advanced-retarded'' representation. The supermatrix $\sigma_{i}$
can be
parametrized
as $\sigma_i=T_{i}^{-1}P_{i}T_{i}$, with $P_i$ being block-diagonal
in the ``advanced-retarded'' decomposition and $T_{i}$ belonging to a
certain graded coset space, see the reviews \cite{efe,vwz}.
For $b\gg 1$ the integral over the matrices $P_i$ can be evaluated
by the saddle-point method \cite{FM}.
Then the action (\ref{4}) is reduced to a $\sigma-$model on a lattice:
\begin{equation}\label{6}
{\cal S}\{Q\}=-\frac{1}{4}(\pi\nu A_0)^2\mbox{Str}
\sum_{ij}\left[(A^{-1})_{ij}-A_0^{-1}\delta_{ij}\right]Q_iQ_j-
\frac{i\pi\nu\omega}{4}\sum_{i}\mbox{Str}{Q_i\Lambda}    \;.
\end{equation}

Here $Q_{i}=T_i^{-1}\Lambda T_i$ satisfies the constraint $Q_i^2=1$,
$A_0$ is given by\footnote{The expression for $A_0$ is valid for
$\alpha>1/2$
and in the limit $N\to \infty$. When $\alpha<1/2$,
$A_0$ starts to depend on $N$
which can be removed by a proper rescaling of the matrix elements
$H_{ij}$
in eq. (\ref{1a}). Then, the properties of the model with
 $\alpha<1/2$ turn out to be equivalent to those of the GOE, so we will not
 consider this case any longer.}
$A_0=\sum_{l}A_{kl}\approx\sum_{r=-\infty}^{\infty}a^2(r)$,
and $\nu$ is the density of states:
\begin{equation}\label{7}
\nu=\frac{1}{2\pi A_0}\left(4A_0-E^2\right)^{1/2}  \;.
\end{equation}

The standard next step is to restrict oneself to the long wavelength
fluctuations of the $Q-$field. For usual RBM characterized by a
function
$a(r)$ decreasing faster than any power of $r$ as $r\to \infty$,
 this is achieved by the momentum expansion of the first term in
the action
 (\ref{6}):
\begin{eqnarray}\nonumber
\sum_{ij}\left[(A^{-1})_{ij}-A_{0}^{-1}\delta_{ij}\right]Q_iQ_j\equiv
\sum_q\left[A_q^{-1}-A_0^{-1}\right]Q_qQ_{-q}\\
\approx \frac{B_2}{2A_0}
\sum_qq^2Q_qQ_{-q}=\frac{B_2}{2A_0}\int dx\left(\partial_x
Q\right)^2   \;,
\label{8}
\end{eqnarray}
where $B_2=\sum_lA_{kl}(k-l)^2$, as defined in eq. (\ref{1}).
This immediately leads to the standard continuous version of
 the nonlinear $\sigma-$model:
\begin{equation}\label{9}
{\cal S}\{Q\}=-\frac{\pi\nu}{4}\mbox{Str}\int dx \left[\frac{1}{2}D_{0}
\left(\partial_xQ\right)^2+i\omega Q\Lambda\right]
\end{equation}
with the classical diffusion constant $D_0=\pi\nu B_2$, which implies
the exponential localization of eigenstates with the localization length
$\xi=\pi\nu D_0\propto b^2$.

Let us try to implement the same procedure for the present case of
power-like bandshape, eq. (\ref{3}). Restricting ourselves to the
lowest  order
term in the momentum expansion, one arrives again at eq. (\ref{9}) as
long as $\alpha\ge 3/2$. This suggests that
for $\alpha\ge 3/2$ the eigenstates
of the present model should be localized in the spatial domain of the
extension $\xi\propto \nu D_0$. However, in contrast to the usual RBM
model we expect
this localization to be power-like rather than exponential:
$|\psi(r)|^2\propto r^{-2\alpha}$ at $r\gg\xi$.
This is quite evident due to the possibility of direct hopping with the same
power-law. On a more formal level the appearance of power-law tails of
wavefunctions is a consequence of the breakdown of the momentum expansion for
the function $A_q^{-1}-A_0^{-1}$ in higher orders in $q^2$.
The presence of power-law ``tails'' of the wave functions,  with an
exponent
$\alpha$ determined by the decay of hopping elements, was found in
numerical simulations in Ref. \cite{oono}.

The most interesting region $1/2<\alpha<3/2$ requires a separate
consideration.
The matter is that eq. (\ref{8}) loses its validity in view of the
divergency
 of the coefficient $B_2$. Instead, a close inspection shows that
\begin{eqnarray}\nonumber
A_0^2(A_q^{-1}-A_0^{-1})&\approx&A_0-A_q=
2\int_0^{\infty}dr a^2(r)(1-\cos{qr})\\
&=&\frac{2}{|q|}\left\{\int_0^{b|q|}dx (1-\cos{x})+(b|q|)^{2\alpha}
\int_{b|q|}^{\infty}\frac{dx}{x^{2\alpha}}(1-\cos{x})\right\}
\nonumber\\
&\approx& c_{\alpha}b^{2\alpha}|q|^{2\alpha-1}\ \ \mbox{for}\
1/2<\alpha<3/2
\ \mbox{and}\  |q|\ll 1/b \;,
\label{10}
\end{eqnarray}
where $c_{\alpha}=2\int_{0}^{\infty}\frac{dx}{x^{2\alpha}}(1-\cos{x})$
is a numerical constant.

The corresponding long wavelength part of the action,
\begin{equation}\label{11}
{\cal S}_0\{Q\}=-\frac{1}{t}\mbox{Str}\int dq|q|^{2\alpha
-1}Q_qQ_{-q} \;,
\end{equation}
can not be reduced to the local-in-space form in the
coordinate representation any longer. Here, $1/t=\frac{1}{4}
(\pi\nu)^2c_{\alpha}b^{2\alpha}\propto b^{2\alpha-1}\gg 1$
plays the role of coupling constant, justifying the
 perturbative/renormalization group treatment of the model described
in the next sections.

Let us mention, that considering the RBM model as a tight-binding
Hamiltonian,
 the corresponding classical motion described by the master equation
on the same $1D$ lattice is superdiffusive for $1/2<\alpha<3/2$:
$\langle |r|\rangle\propto t^{1/(2\alpha-1)}$. As will be discussed in
Sec. \ref{s6},
this influences the asymptotic behavior of the spectral
correlation function
for the corresponding quantum system.

\section{Perturbative treatment of the non-linear $\sigma-$model:
General formulas.}
\label{s3}

In this section, we derive one-loop perturbative corrections to the
density--density correlation function and inverse participation ratios.
The analysis of these expressions for various values of the power-law
parameter
$\alpha$ will be presented in Sec. \ref{s5}.

\subsection{Density--density correlation function.}
\label{3.1}

The basic object characterizing the behavior of a particle in a
random medium
is the density--density correlation function, which can be generally
expressed in terms of the $\sigma-$model as follows \cite{efe}:
\begin{equation}
K(r_1,r_2;\omega)=-(\pi\nu)^2\int DQ \:Q_{12,\alpha\beta} k_{\beta\beta}
Q_{21,\beta\alpha}e^{-S\{Q\}}  \;.
\label{12}
\end{equation}
Here the indices $p,p'$ of the matrix $Q_{pp',\alpha\beta}$ correspond to
its advanced--retarded block structure, whereas $\alpha,\beta$
discriminate
between bosonic and fermionic degrees of freedom. The matrix
$k_{\beta\beta}$
is equal to 1 for bosons and (-1) for fermions. In order to calculate the
correlation function (\ref{12}) perturbatively, one needs to parametrize
the matrix $Q$ in terms of the independent degrees of freedom. We find it
convenient to use the following parametrization \cite{efe}:
\begin{equation}
Q=\Lambda(W+\sqrt{1+W^2})=\Lambda\left(1+W+{W^2\over 2}-{W^4\over
8}\right)\ ,
\label{13}
\end{equation}
where $W$ is block-off-diagonal in the advanced--retarded representation.
To get the perturbative expansion for $K(r_1,r_2;\omega)$, one has to
substitute eq. (\ref{13}) into (\ref{12}), to separate the part
quadratic in $W$
 from the rest in the exponent and to apply the Wick theorem
(see Ref. \cite{FM1} for the contraction rules). In the usual case, when
the action is given by eq. (\ref{9}), the leading order (tree level)
result
reads in  momentum space as follows:
\begin{equation}
K_0(q,\omega)={2\pi\nu\over D_0 q^2-i\omega} \;.
\label{14}
\end{equation}
The perturbative quantum corrections do not modify the general
form (\ref{14}), but change the value of the diffusion constant. In
particular, in one-loop order one gets eq. (\ref{14}) with $D_0$
replaced by
\cite{wl}
\begin{equation}
D=D_0\left\{1-{1\over\pi\nu V}\sum_{q_i=\pi n_i/L_i}{1\over D_0
q^2-i\omega}
\right\}                       \;.
\label{15}
\end{equation}
This induces the standard weak-localization correction to the
conductivity.

Now we implement an analogous procedure for the non-local $\sigma-$model
of the type of eq. (\ref{11}):
\begin{equation}\label{16}
{\cal S}\{Q\}=\frac{1}{t}\mbox{Str}\sum_{r.r'}U(r-r')Q(r)Q(r')-
i\frac{\pi\nu}{4}\sum_{r}\mbox{Str}{\Lambda Q(r)}\ ,
\end{equation}
with the Fourier-transform of $U(r)$ behaving at small momenta as
\begin{equation}\label{17}
\tilde{U}(q)=-|q|^{\sigma};\quad 1/2<\sigma<2   \;.
\end{equation}
The exponent $\sigma$ is related to the parameter $\alpha$ of the RBM
model by $\sigma=2\alpha-1$. In  leading order,
we keep in the action the terms quadratic in $W$ only, which yields:
\begin{equation}\label{18}
K_0(q,\omega)=\frac{2\pi\nu}{8(\pi\nu t)^{-1}|q|^{\sigma}-i\omega} \;,
\end{equation}
corresponding to a superdiffusive behavior.

To calculate the one-loop correction to $K_0(q)$ (we set $\omega=0$
for simplicity) we expand the kinetic term in ${\cal S}\{Q\}$ up to
 fourth order in $W$:
\begin{equation}\label{19}
\sum_{r,r'}\mbox{Str}\,U(r-r')Q(r)Q(r')\mid_{\mbox{4th order}}
=\sum_{r,r'}\frac{1}{4}
\mbox{Str}\,W^2(r)W^2(r')     \;.
\end{equation}
The contraction rules are given by eqs. (8), (16) of  Ref. \cite{FM1},
with the propagator $\Pi(q)$ replaced by
\begin{equation}\label{20}
\Pi(q)=\frac{t}{8|q|^{\sigma}}   \;.
\end{equation}

There is only one  one-loop diagram contributing to the self-energy part
in the present parametrization, see Fig. 1. Evaluating it, we find:
\begin{equation}\label{21}
\delta
\Gamma_1^{(2)}=\frac{1}{2}\int(dk)\frac{|q+k|^{\sigma}}{|k|^{\sigma}} \;,
\end{equation}
where $\int (dk)\equiv\int\frac{dk}{2\pi}\equiv N\sum_{k}$. There is also
a contribution to $ \Gamma^{(2)}$
from the Jacobian of transformation (\ref{13}),
which is equal to \cite{ZJ}
\begin{equation}\label{22}
\delta\Gamma_2^{(2)}=-\frac{1}{2}\delta(0)\equiv -\frac{1}{2}\int (dk) \;.
\end{equation}
Combining eqs. (\ref{21}) and (\ref{22}), we get
\begin{equation}\label{23}
\delta\Gamma^{(2)}=\frac{1}{2}\int(dk)\frac{|q+k|^{\sigma}-|k|^{\sigma}}
{|k|^{\sigma}} \;.
\end{equation}

Finally, we get the following expression for the density-density
correlation
function up to one-loop order:
\begin{equation}
K^{-1}(q)=K_0^{-1}(q)-\frac{(\pi\nu)^2}{2}
\int(dk)\frac{|q+k|^{\sigma}-|k|^{\sigma}}
{|k|^{\sigma}}    \;.
\label{24}
\end{equation}

\subsection{Inverse participation ratio.}
\label{s3.2}
In order to characterize eigenfunctions statistics quantitatively, it is
convenient to introduce a set of moments
$I_q=\sum|\psi(r)|^{2q}$ of the eigenfunction local intensity
$|\psi(r)|^2$. The quantity $I_2$ is known as the inverse participation
ratio (IPR). The whole set of moments $I_q$ is a useful measure of the
eigenfunction structure. For completely ``ergodic'' eigenfunctions
covering randomly but uniformly the whole sample,
$\langle I_q\rangle ={(2q-1)!!\over
N^{q-1}}$, as in GOE. In contrast, if eigenfunctions are localized in
a domain of  size $\xi$, then $\langle I_q\rangle\propto
1/\xi^{q-1}$.
Finally,
the multifractal structure of the wavefunction manifests itself via
the dependence $\langle I_q\rangle\propto 1/N^{d_q(q-1)}$
with $d_q<1$ being the set
of fractal dimensions \cite{weg,cast}.

The method allowing one to calculate perturbative corrections to
the GOE-like
results in the weak localization regime was developed in \cite{KM,FM1}. It
is straightforwardly applicable to the present case of power-law RBM,
provided the appropriate modification of the diffusion propagator
entering the contraction rules is made,
see the text preceding eq. (\ref{20}). One finds
\begin{equation}
\langle I_q\rangle
=\left\{1+{1\over N}q(q-1)\sum_r\Pi(r,r)\right\}{(2q-1)!!\over
N^{q-1}}\ ,
\label{25}
\end{equation}
 where $\Pi(r,r')=(1/N)\sum_q\Pi(q)\exp[iq(r-r')]$ and $\Pi(q)$ is
given by eq. (\ref{20}).

\section{Renormalization group treatment.}
\label{s4}

Our effective $\sigma-$model, eq. (\ref{16}), is actually of
one-dimensional
nature. However, for the sake of generality,  in the
present section we find it convenient
to consider it to be defined in $d$-dimensional space
with arbitrary $d$. The form (\ref{18}) of the generalized diffusion
propagator suggests that $d=\sigma$ should play the role of the logarithmic
dimension for the problem. In the vicinity of this critical value it is
natural to carry out a renormalization group (RG) treatment of the
model.
We will follow the procedure developed for general non-linear
$\sigma-$models
in \cite{ZJ}. We start from expressing the action in terms of the
renormalized coupling constant $t=Z_1^{-1}t_B\mu^{d-\sigma}$, where $t_B$
is the bare coupling constant and $\mu^{-1}$ is the length scale
governing
the renormalization \footnote{Note that the $W$-field renormalization
is absent due to the supersymmetric character of the problem, which is
physically related to the particle number conservation
\protect\cite{efe}.}:
\begin{eqnarray}
S&=&{\mu^{d-\sigma}\over 2t Z_1}\sum_{rr'}U(|r-r'|)\:\mbox{Str}\,
\left[-W(r)W(r')+\sqrt{1+W^2(r)}\sqrt{1+W^2(r')}\right]
\nonumber \\    &-&{i\pi\nu\omega\over 4}
\sum_r \mbox{Str}\sqrt{1+W^2(r)}         \;.
\label{26}
\end{eqnarray}
Expanding the action in powers of $W(r)$ and keeping  terms up to
4-th order, we get
\begin{eqnarray}
S&=&S_0+S_1+O(W^6) \;, \nonumber\\
S_0&=&{\mu^{d-\sigma}\over 4t Z_1}\sum_{rr'}U(|r-r'|)\:\mbox{Str}\,
(W(r)-W(r'))^2-{i\pi\nu\omega\over 8}\sum_r \mbox{Str}\,W^2(r) \;,
\nonumber\\
S_1&=&{\mu^{d-\sigma}\over 8t Z_1}\sum_{rr'}U(|r-r'|)\mbox{Str}\,
W^2(r)W^2(r')+{i\pi\nu\omega\over 32}\sum_r \mbox{Str}\,W^4(r) \;.
\label{27}
\end{eqnarray}
We have restricted ourselves to 4-th order terms, since they are
sufficient for obtaining the renormalized quadratic part of the action
in one-loop order. The calculation yields, after the cancellation of an
$\int(dk)\propto \delta(0)$ term with the contribution of the Jacobian:
\begin{equation}
S_{{\rm quad}}=S_0+\langle S_1\rangle={1\over 2}\sum_q\mbox{Str}W_q W_{-q}
\left[-{\mu^{d-\sigma}\over t Z_1}\tilde{U}(q)-{1\over 2}
\int(dk){\tilde{U}(k)-\tilde{U}(k+q)\over
-{\mu^{d-\sigma}\over t Z_1}\tilde{U}(k)-{i\pi\nu\omega\over 4}}\right]
\;.
\label{28}
\end{equation}
According to the renormalization group idea, one has to chose the
constant
$Z_1(t)=1+at+\ldots$ so as to cancel the divergency in the
coefficient in front of the leading $|q|^\sigma$ term. This will be
done in
the next section.

\section{Analysis of the model for different values of the exponent
$\sigma=2\alpha-1$.}
\label{s5}

\subsection{$1<\sigma<2$ ($1<\alpha<3/2)$.}
\label{s5.1}

To evaluate the one-loop correction (\ref{24}) to the diffusion
propagator,
we use the expansion
\begin{equation}\label{29}
{|\bbox{q}+\bbox{k}|^\sigma\over|\bbox{k}|^\sigma}-1\simeq\left\{
\begin{array}{ll}
\displaystyle{
\sigma{\bbox{qk}\over k^2}+{\sigma\over 2}{q^2\over k^2}+
\sigma\left({\sigma\over 2}-1\right)\left({\bbox{qk}\over k^2}\right)^2
+\ldots   }
\ ,&\qquad q\ll k\\
\displaystyle{
{|q|^\sigma\over |k|^\sigma}     }
\ ,&\qquad q\gg k
\end{array}\right.  \;.
\end{equation}
Thus, the integral in eq. (\ref{24}) can be estimated as
\begin{equation}
I\equiv\int(dk)\left[
{|\bbox{q}+\bbox{k}|^\sigma\over|\bbox{k}|^\sigma}-1\right]
\approx\int_{k>q}(dk)\sigma\left(1+{\sigma-2\over
d}\right){q^2\over 2k^2}
+\int_{k<q} (dk) {|q|^\sigma\over |k|^\sigma}  \;.
\label{30}
\end{equation}
For $\sigma>d=1$ the integral diverges at low $k$, and  the second
term in eq. (\ref{30})  dominates. This gives
\begin{equation}
I\sim L^{\sigma-1}|q|^{\sigma}\ ,
\label{31}
\end{equation}
where $L$ is the system size determining the infrared cut-off (in the
original RBM formulation it is just the matrix size $N$). This leads
to the following one-loop expression for the diffusion correlator
\begin{equation}
K(q)={(\pi\nu)^2\tilde{t}\over 4|q|^\sigma}\ ; \qquad
\tilde{t}^{-1}=t^{-1}-\mbox{const} \, L^{\sigma-1}  \;.
\label{32}
\end{equation}

Now we turn to the renormalization group analysis, as described in
Sec. \ref{s4}. In one-loop order, the expression for the
renormalization
constant $Z_1$ is determined essentially by the same integral $I$,
eq. (\ref{30}), with the RG scale $\mu$ playing the
role of the infrared cut-off (analogous to that of the system size $L$
in eq. (\ref{32})). This yields (in the minimal subtraction scheme):
\begin{equation}
Z_1(t)=1-{1\over 2\pi}{t\over\sigma-1}+O(t^2)   \;.
\label{33}
\end{equation}
It is easy to check that this results in a relation between the bare and
the renormalized coupling constant analogous to eq. (\ref{32}):
\begin{equation}
{1\over t}\mu^{1-\sigma}\equiv{Z_1\over t_B}={1\over t_B}-
{1\over 2}{1\over\sigma-1}\mu^{1-\sigma}  \;.
\label{34}
\end{equation}
~From eq. (\ref{33}), we get the expression for the $\beta$-function:
\begin{equation}
\beta(t)=\left.{\partial t\over\partial\ln\mu}\right|_{t_B}=
{(1-\sigma)t\over 1+t\partial_t\ln Z_1(t)}=-(\sigma-1)t-{t^2\over 2\pi}+
O(t^3)  \;.
\label{35}
\end{equation}
Both eqs. (\ref{32}) and (\ref{35}) show that the coupling constant $t$
increases with the system size $L$ (resp. scale $\mu$), which is
analogous
to the behavior found in the conventional scaling theory of localization
in $d<2$ dimensions \cite{AALR,weg1}. The RG flow reaches the strong
coupling regime $t\sim 1$ at the scale $\mu\sim t_B^{1/(\sigma-1)}$.
Remembering the relation of the bare coupling constant $t_B$ and the
index $\sigma$ to the parameters of the original PRBM model:
$t_B^{-1}\propto b^{2\alpha-1}$; $\sigma\propto 2\alpha-1$, we conclude
that the length scale
\begin{equation} \label{36}
\xi\sim t_B^{-1/(\sigma-1)}\sim b^{2\alpha-1\over 2\alpha-2}
\end{equation}
plays the role of the
localization length for the PRBM model.

This conclusion is also supported by an inspection of the
expression for the
IPR, eq. (\ref{25}). Evaluating the one-loop perturbative correction in
eq. (\ref{25}), one gets
\begin{equation}
\langle I_q\rangle
={(2q-1)!!\over N^{q-1}}\left\{1+q(q-1){t\over 8\pi^\sigma}\zeta(\sigma)
N^{\sigma-1}\right\}\ ,
\label{37}
\end{equation}
where $\zeta(\sigma)$ is Riemann's zeta-function;
$\zeta(\sigma)\simeq
1/(\sigma-1)$ for $\sigma$ close to unity. The correction term becomes
comparable to the leading (GOE) contribution for the system size
$N\sim t^{1/(\sigma-1)}$,
parametrically coinciding with the localization length $\xi$. For larger
$N$ the perturbative expression (\ref{37}) loses its validity, and the IPR
is expected to saturate at a constant value
$\langle I_q\rangle\sim\xi^{1-q}$ for
$N\gg\xi$.

In conclusion of this subsection, let us stress once more that the
localized
eigenstates in the present model are expected to have integrable
power-law
tails: $|\psi^2(r)|\propto r^{-2\alpha} = r^{-\sigma-1}$ at $r\gg\xi$.

\subsection{$0<\sigma<1\ \ (1/2<\alpha<1)$}
\label{s5.2}
We start again from considering the perturbative corrections
to the diffusion propagator (\ref{24}). The integral (\ref{30})
is now dominated by the region $k\sim q$, and is proportional
to $|q|$. We get, therefore:
\begin{equation}\label{39}
(\pi\nu)^2 K^{-1}(q)=\frac{4|q|^{\sigma}}{t}-C_{\sigma}|q|
\end{equation}
with a numerical constant $C_{\sigma}$.

We see that the correction term is of higher order in
$|q|$ as compared to the leading one.
Thus, it does not lead to a renormalization of the coupling constant $t$.
This is readily seen also in the framework of the RG scheme,
where the one-loop integral in eq. (\ref{28}) gives no
rise to  terms of the form $|q|^{\sigma}$.
One can check that this feature is not specific to the
one-loop RG calculation, but holds in higher orders as well.
An analogous conclusion was reached for the case of a vector model
with long-ranged interaction by Brezin et al. \cite{bre}.

In our case this implies that the renormalization constant $Z_1$
is equal to unity and the $\beta-$function is trivial:
\begin{equation}\label{40}
\beta(t)=(1-\sigma)t    \;.
\end{equation}
This means that the model does not possess a critical
 point, and, for $\sigma<1$, all states  are delocalized, for any value
of the bare coupling constant $t$. This property should be contrasted
with the behavior of a $d-$dimensional conductor
 described by the conventional local non-linear $\sigma-$
model and undergoing an Anderson transition at some critical coupling
$t=t_c$ \cite{weg1}.

Since all states of the model turn out to be delocalized, one
 might think that their statistical properties are the same as for GOE.
This is, however, not quite true. In particular, calculating the
{\it variance}
of the inverse participation ratio $I_2$ in the same way as it is done in
Ref. \cite{FM1} we get:
\begin{equation}\label{41}
\delta(I)=\frac{\langle I_2^2\rangle-\langle I_2\rangle ^2}
{\langle I_2\rangle ^2}=
\frac{8}{N^2}\sum_{rr'}\Pi^2(r,r')=\frac{8}{N^2}\sum_{q=\pi n/N;\:
n=1,2,..}
\Pi^2(q) \;.
\end{equation}
At $\sigma>1/2$ the sum over $q$ is convergent yielding
\begin{equation}\label{42}
\delta(I)=\frac{t^2}{8\pi^{2\sigma}}\frac{\zeta(2\sigma)}{N^{2-2\sigma}}
\;.
\end{equation}
Thus, in this regime the fluctuations of the IPR are much stronger
than for the GOE where $\delta(I)\propto 1/N$. Only for $\sigma < 1/2$
 ($\alpha < 3/4$)
the IPR fluctuations acquire the GOE character.

Considering higher cumulants of the IPR, $\langle\!\langle
I_2^n\rangle\!\rangle$,
one finds that the GOE behavior is restored at
$\sigma<\sigma_c^{(n)}\equiv 1/n$. In this sense, the model is
analogous to a
 $d$-dimensional conductor at $d=2/\sigma$.
 Therefore, only when $\sigma\to 0$
 (correspondingly, $\alpha\to 1/2$ in the original PRBM formulation)
all statistical properties become equivalent to those typical for the GOE.

\subsection{Critical regime: $\sigma=1$ ($\alpha=1$).}
\label{s5.3}

As we have seen, the case $\sigma=1$ separates the regions of
localized ($\sigma>1$) and extended ($\sigma<1$) states. It is then
natural
to expect some critical properties showing up just at $\sigma=1$.
Let us again start from considering the generalized diffusion propagator,
 eq. (\ref{24}). At $\sigma=1$ the one-loop correction yields
\begin{equation}\label{43}
\frac{1}{(\pi\nu)^2}K^{-1}(q)=4|q|\left[t^{-1}-\frac{1}{2}\ln(|q|L)\right]
\;.
\end{equation}
As it was natural to expect for the critical point, the correction to the
coupling constant is of logarithmic nature. However, eq. (\ref{43})
differs essentially from that typical for a $2D$ disordered conductor:
\begin{equation}\label{44}
t^{-1}=t_B^{-1}-\ln(L/l)\ ,
\end{equation}
where the bare coupling constant $t_B$ corresponds to  scale $l$.
Comparing the two formulas, we see that in eq. (\ref{43}) the mean
free path $l$ is replaced by the inverse momentum $q^{-1}$. Therefore,
the correction to the bare coupling constant is small for low momenta
$q\sim 1/L$, and the correlator $K(q)$ is not renormalized.
This implies the absence of eigenstate localization, in contrast
 to the $2D$ diffusive conductor case, where eq. (\ref{44}) results in an
exponentially large localization length $\xi\propto \exp{t_B^{-1}}$.
On a more formal level, the absence of essential corrections to
the low-$q$ behavior of $K(q)$ is due to the fact that the region $k>q$
does not give a logarithmic contribution.
This is intimately connected with the absence of $t$
 renormalization at $\sigma<1$.

To study in more details the structure of critical eigenfunctions,
let us consider the set of IPR, $I_q$. The perturbative correction,
 eq. (\ref{25}), is evaluated at $\sigma=1$ as
\begin{equation}\label{45}
\langle I_q\rangle
=\left\{1+q(q-1)\frac{t}{8\pi}\ln{(N/b)}\right\}\frac{(2q-1)!!}
{N^{q-1}}    \;,
\end{equation}
where the microscopic
scale $b$, eq. (\ref{3}), enters as the ultraviolet cut-off
for the $\sigma-$model, the role usually played by the mean free
path $l$.
This formula is valid as long as the correction is small: $q\ll
\left[\frac{t}{8\pi}\ln{(N/b)}\right]^{-1/2}$. For larger $q$
the perturbation theory breaks down, and one has to
 use the renormalization group approach. This procedure, pioneered
by Wegner
\cite{weg} and developed by Altshuler, Kravtsov and Lerner \cite{AKL}
requires the introduction of higher vertices of the type $z_q\int
\mbox{Str}^q
(Qk\Lambda)dr$ in the action of the non-linear $\sigma-$model, and their
subsequent renormalization. This  results in  RG equations
for the charges $z_q$ which in the present case, and in  one-loop order,
read:
\begin{equation}\label{46}
\frac{dz_q}{d\ln{\mu^{-1}}}=q(q-1)\frac{t}{8\pi}z_q  \;,
\end{equation}
where $\mu^{-1}$ is the renormalization scale. Integrating eq. (\ref{46}),
we find:
\begin{equation}\label{47}
\langle I_q\rangle
=\frac{(2q-1)!!}{N^{q-1}}{\left(N\over b\right)}^{q(q-1)\frac{t}{8\pi}} \;.
\end{equation}
Note, that this formula is reduced to the perturbative expression,
eq. (\ref{45}), in the regime
$q\ll\left[\frac{t}{8\pi}\ln{(N/b)}\right]^{-1/2}$.

The behavior described by eq. (\ref{47}) is characteristic for
 a multifractal structure of wave functions, when
\begin{equation}\label{48}
\langle I_q\rangle\propto N^{-d_q(q-1)}
\end{equation}
with $d_q$ being the set of fractal dimensions. We find from eq. (\ref{47}):
\begin{equation}\label{49}
d_q=1-q\frac{t}{8\pi} \;.
\end{equation}
This form of the fractal dimensions is similar to that found in two
and $2+\epsilon$ dimensions for the usual diffusive conductor
\cite{weg,cast,AKL}.
The one-loop result (\ref{49}) holds for $q\ll 8\pi/t$.

The set of fractal dimensions $d_q$ (as well as spectral properties at
 $\sigma=1$, see next section) is parametrized by the coupling
constant $t$.
Strictly speaking, our $\sigma-$model derivation is justified for
$t\ll 1$
 (i. e. $ b\gg 1$). However, the opposite limiting case can be also
studied,
following the ideas of Levitov \cite{levitov}. This corresponds to
a $d-$dimensional Anderson insulator, perturbed by a weak long-range
 hopping with an amplitude decreasing with distance as $r^{-\sigma}$.
The arguments of Levitov \cite{levitov} suggest that the states
delocalize at $\sigma\le d$, carrying some fractal properties at
$\sigma=d$.
Our PRBM model in the limit $b\ll 1$ is just the $1D$ version of
this problem.
This shows that the conclusion about localization (delocalization)
of eigenstates for $\sigma>1$ (resp. $\sigma<1$), with $\sigma=1$
being a critical point holds irrespective of the particular value of the
parameter $b$. Alternatively, the regime of the Anderson insulator
with weak
power-law hopping can be described in the framework of the non-linear
 $\sigma-$model, eq. (\ref{16}), by considering the limit $t\gg 1$.
Formally, the non-linear $\sigma-$model for arbitrary $t$ can be
derived from a microscopic tight-binding model by allowing $n\gg 1$
``orbitals'' per site \cite{weg1}.

At any rate, the PRBM model, eqs. (\ref{1a}), (\ref{3}), with
 arbitrary $0<b<\infty$, or the $\sigma-$model, eq. (\ref{16}), with
arbitrary
 coupling constant $0<t<\infty$ display at $\sigma=1$ a
 rich critical behavior parametrized by the value of $b$
(respectively, $t$).

\section{Spectral properties.}
\label{s6}

Let us consider now the issue of spectral statistics of the PRBM
model. As
is well known, a usual diffusive conductor exhibits the
Wigner--Dyson (RMT)
level statistics in the limit of infinite dimensionless conductance
$g=2\pi\nu DL^{d-2}$. At finite $g\gg 1$, there appear deviations
\cite{AS,KM,AA}. In the present section we would like to address the
analogous problem in the case of the PRBM model.

The basic quantity characterizing the spectral properties is the
two-level
 correlation function
\begin{equation}
R(s)={1\over\langle\nu\rangle^2}\langle\nu(E)\nu(E+\omega)\rangle\ ,
\label{50}
\end{equation}
where $s=\omega/\Delta$, $\Delta$ is the mean level spacing,
$\nu(E)$ is the
density of states at  energy $E$, and $\langle\ldots\rangle$ denotes
the ensemble averaging. Following \cite{KM}, we find the leading
correction
to the Wigner--Dyson form $R^{WD}(s)$ of the level correlation function,
eq. (\ref{50}), as
\begin{equation}
R(s)=\left[1+{1\over 2}{\cal C}{d^2\over ds^2}s^2\right]R^{WD}(s)\ ,
\label{51}
\end{equation}
where
\begin{equation}
{\cal C}={1\over N^2}\sum_{rr'}\Pi^2(r,r')={1\over N^2}
\sum_{q=\pi n/N;\:n=1,2,\ldots}
\Pi^2(q)=\left\{
\begin{array}{ll}
\displaystyle{t^2\over 64\pi^{2\sigma}}N^{2\sigma-2}\ ,& \qquad
\sigma>1/2\\
\mbox{const}
\displaystyle{t^2\over 64\pi^{2\sigma}b^{1-2\sigma}}N^{-1}\ ,& \qquad
\sigma<1/2\ .
\end{array}
\right.
\label{52}
\end{equation}
At $\sigma<1/2$ the sum divergent at high momenta  is cut off at
$q\sim\pi/b$;
the procedure leaving undetermined a constant of order of unity.

The correlation function $R(s)$ is close to its RMT value if $\sigma<1$
(the region of delocalized states), or else if $\sigma>1$ and the system
size $N$ is much less than the localization length $\xi$, eq. (\ref{36}).
Under these conditions, eq. (\ref{51}) holds as long as the
correction term
is small compared to the leading one. This requirement produces the
following restriction on the frequency $s=\omega/\Delta$:
\begin{equation}
s<s_c\sim\left\{
\begin{array}{ll}
t^{-1}N^{1-\sigma}\ , &\qquad \sigma>1/2\\
t^{-1}b^{{1\over 2}-\sigma}N^{1/2}\propto (Nb)^{1/2}\ , &\qquad
\sigma<1/2
\end{array}
\right. \;.
\label{53}
\end{equation}
At larger frequencies ($s>s_c$), the form of the level correlation
function
changes from the $1/s^2$ behavior typical for RMT to a completely
different
one \cite{AS}. Extending the calculation of Ref. \cite{AS} to the present
case, we find
\begin{eqnarray}
R(s)&=&{\Delta^2\over \pi^2}\mbox{Re}\sum_{n=0,1,...}
{1\over \left[{8\over\pi\nu t}\left({\pi n\over
N}\right)^\sigma-i\omega\right]
^2}\nonumber\\ \label{54} &\propto&
\left\{
\begin{array}{ll}
N^{1-1/\sigma}t^{1/\sigma}s^{1/\sigma-2}\ ,&\qquad \sigma>1/2\ \
(\sigma\ne 1)\\
t^2N^{-1}b^{2\sigma-1}\propto (Nb)^{-1}\ ,&\qquad \sigma<1/2
\end{array}
\right.     \;.
\end{eqnarray}

At last, let us consider the level statistics in the critical regime
$\sigma=1$. In this case the coefficient of proportionality  in the
asymptotic
expression (\ref{54}) vanishes in view of  analyticity:
\begin{equation}
R(s)\sim {\Delta t\over
16\pi^2}\int_{-\infty}^{\infty}{dx\over(x-i\omega)^2}
=0                  \;.
\label{55}
\end{equation}
This is similar to what is known to happen in the case of a $2D$ diffusive
conductor \cite{altland,kravtsov}. A more accurate consideration
requires
taking into account the high-momentum cut-off at $q\sim b^{-1}$.
In full analogy with the  $2D$ situation mentioned
\cite{altland,AM,kravtsov},
we find then a linear term in the level number variance:
\begin{eqnarray}
&& \langle\delta N^2(E)\rangle\simeq\kappa\langle N(E)\rangle\nonumber\\
&& \kappa=\int R(s)ds={t\over 8\pi}                \;.
\label{56}
\end{eqnarray}
The presence of the linear term (\ref{56}) (as well as the
multifractality
of eigenfunctions, Sec. \ref{s5}) makes the case $\sigma=1$  similar
to the
situation on the mobility edge of a disordered conductor in $d>2$
\cite{AM}.
Let us finally mention that the value of $\kappa$, eq. (\ref{56}), is in
agreement with the formula $\kappa=(d-d_2)/2d$, suggested recently by
Kravtsov \cite{krav}, where $d_2$ is given by eq. (\ref{49}) with $q=2$,
and $d=1$ in the present case.

\section{Numerical simulations}
\label{s7}

We have performed numerical simulations of the PRBM model  for
values of
$\alpha\in [0,2]$ and $b=1$. In Fig. 2 we present typical
eigenfunctions for
 four different regions of $\alpha$. In agreement with the
theoretical picture
presented above, the eigenstates corresponding to $\alpha=0.375$ and
$\alpha=0.875$ are extended, whereas those corresponding to $\alpha=1.25$
and $\alpha=1.625$ are localized. At the same time, one can notice
that the
states with $\alpha=0.875$ and $\alpha=1.25$ exhibit a quite sparse
structure,
as opposed to the other two cases. We believe that this can be explained
by the proximity of the former two values of $\alpha$ to the
critical value
$\alpha=1.0$, where eigenstates should show the multifractal
behavior, see
Sec. \ref{s5}.

In order to get a more quantitative insight into the properties of the
eigenstates, we concentrated our attention on the behavior of the mean value
of the IPR, $\langle I_2\rangle$, and on the relative variance,
$\delta=(\langle I_2^2\rangle - \langle I_2\rangle^2)/\langle
I_2\rangle^2$.
At any given $\alpha$ we studied the dependence of the quantities
$\langle I_2\rangle$ and $\delta$ on the matrix size $N$ and approximated
these dependencies by the power-laws $\langle I_2\rangle\propto
1/N^\nu$,
$\delta\propto 1/N^\mu$ for $N$ ranging from $100$ to $2400$.
On Figs. 3 and 4 we plotted the values of the exponents
$\nu$ and $\mu$ obtained
in this way, versus the PRBM parameter $\alpha$. The expected theoretical
curves following from the results of Sec. \ref{s5} are presented as well.
We see from Fig. 3 that the data show a crossover  from the behavior
typical for extended states ($\nu=1$) to that typical for localized
states
($\nu=0$), centered approximately at the critical point $\alpha=1$. We
attribute the deviations from the sharp step-like theoretical curve
$\nu(\alpha)$ to the finite-size effects which are unusually pronounced
in the PRBM model due to the long-range nature of the off-diagonal
coupling.
The data for the exponent $\mu$ (Fig. 4)
also show a reasonable agreement with the
expected linear crossover, $\mu=4(1-\alpha)$ for $3/4<\alpha<1$, see
eq. (\ref{42}).

\section{Conclusion.}
\label{s8}

In this paper, we have performed a detailed investigation of the
RBM model with a power-law decay of the matrix elements, eq. (\ref{3}).
Physically, one can look at this model as describing a particle in a $1D$
disordered system with a power-law hopping term. As a theoretical tool,
we have used a mapping of the problem onto a supermatrix non-linear
$\sigma-$model, eqs. (\ref{16}), (\ref{17}). Depending on  the value of
the power-law exponent $\alpha$ in eq. (\ref{3}) (or, equivalently,
$\sigma=2\alpha-1$ in eq. (\ref{17})), three different regimes are found:
For $\alpha>1$ ($\sigma>1$) all eigenstates are localized with integrable
power-law tails. For $\alpha<1$ ($\sigma<1$) the eigenstates are
delocalized,
for any value of the bandwidth $b$ of the PRBM model, eq. (\ref{3})
(resp. coupling constant $t\propto b^{1-2\alpha}$ of the $\sigma-$model,
eq. (\ref{17})). These two regimes are separated by the critical value
$\alpha=1$ ($\sigma=1$), where the structure of eigenstates is
multifractal,
and energy levels show statistics intermediate between Wigner--Dyson
and Poisson ones. These critical properties are similar to those found
on the mobility edge of a $d$-dimensional disordered conductor.
At $\sigma=1$, we find a family of  critical points labeled by the
value of the coupling constant $t$ of the non-linear $\sigma-$model,
so that the critical behavior is parametrized by the value of $t$.
In particular, it determines the multifractal exponents $d_q$ and the
coefficient $\kappa$ of the linear term in the level number
variance, which are
given at $t\ll 1$ by eqs. (\ref{49}) and (\ref{56}), respectively.

Turning our attention to the regime of localized states, we find that
it can be subdivided into two domains. In the region $\alpha>3/2$ the
properties of the model are rather close to those of a conventional
quasi-$1D$ conductor \cite{RBMrev}. On the other hand, for $1<\alpha<3/2$
the wave packet spreading on a short time scale is superdiffusive,
$\langle|r|\rangle\sim t^{1/(2\alpha-1)}$, which leads to a modification
of the Altshuler--Shklovskii ``tail'' of the spectral correlation function,
eq. (\ref{54}), and to an unusual scaling of the localization length
$\xi\propto b^{2\alpha-1}$. The regime of extended states, $\alpha<1$,
can be also subdivided into two domains. For $\alpha<1/2$, all
statistical
 properties are identical to those of the GOE (which corresponds to
$\alpha=0$).
On the other hand, at $1/2<\alpha<1$ the model is quite similar to
a diffusive
conductor in $d=(\alpha-1/2)^{-1}$ dimensions. This is reflected,
in particular, in the fluctuations of the IPR (see eq. (\ref{42}) and
the discussion
following it), and in the large-frequency ``tail'' of the spectral
correlator, eq. (\ref{54}).

Our conclusion about the existence of a set of critical theories
at $\alpha=1$, parametrized by
 the coupling constant $t$, perfectly agrees with earlier results
by Levitov \cite{levitov}. He studied the effect of a weak power-law
hopping on an Anderson insulator, which corresponds to the limit $t\gg 1$
in the $\sigma-$model formulation, and arrived at the conclusion of
criticality of the model at $\alpha$ equal to the spatial dimension $d$.
Let us also note that our results are in accordance with the fact
of absence
of localization effects in the Quantum Fermi Accelerator model
\cite{jose}.
As it was pointed out in \cite{jose}, the evolution equation for
this model
takes the form of a finite difference equation of the tight binding
type with
a long range hopping term decaying in a power-law fashion,
eq. (\ref{3}), with
$\alpha=1$. Our results show that this case corresponds to the critical
point with extended eigenstates and intermediate level statistics,
in  agreement with the behavior found in Ref. \cite{jose}.

We have presented results of a direct numerical simulation of the
PRBM model.
Our data are in reasonable agreement with the above theoretical picture.
However, a more detailed numerical investigation of the structure of
eigenstates and of spectral statistics is certainly
desirable. In particular, it would
be very interesting to study the critical manifold, $\alpha=1$, where
the multifractal properties of eigenstates and intermediate level
statistics
are predicted by our theory.

\section{Acknowledgments.}

We are grateful to F. Izrailev and I. Guarneri for useful discussions
on early stage of this work, and to V. E. Kravtsov, M. R. Zirnbauer,
B. L. Altshuler, and A. M. Tsvelik for
discussion of the results. A.D.M. and Y.V.F. acknowledge with thanks
the warm hospitality extended to them during the Program ``Quantum
Chaos in Mesoscopic Systems'' in the Institute for Theoretical Physics
in Santa Barbara, where this research was completed. This research
was supported in part by the National Science Foundation under Grant No.
PHY94-07194 (A.D.M. and Y.V.F.), the Deutsche
Forschungsgemeinschaft by SFB 195 (A.D.M.) and SFB 237
``Unordnung und Gro\ss{}e Fluktuationen''
(Y.V.F.), the Minerva Foundation (F.-M.D.), and the Minerva Center for
Nonlinear Physics of Complex Systems (Y.V.F. and F.-M.D.).

\newpage

\noindent
{\Large {\bf Figure captions}}\\[1cm]

\noindent
{\bf Fig. 1} \\
One-loop diagram contributing to the self-energy part of the density-density
correlation function. The solid line denotes the $Q$-field propagator,
whereas the dashed line represents the interaction $U(r-r')$. 
\\[3mm]

\noindent
{\bf Fig. 2} \\
Typical eigenfunctions for the matrix size $N=800$ and
 four different values of $\alpha$: a)$\alpha=0.375$;
b)$\alpha=0.875$; c)$\alpha=1.250$; d)$\alpha=1.625$.
\\[3mm]

\noindent
{\bf Fig. 3} \\
Index $\nu $ characterizing the dependence of the inverse participation
ratio $\langle I_2\rangle$ on the matrix size $N$ via
$\langle I_2\rangle\propto 1/N^\nu$, as a function of $\alpha $. Points
refer to the best-fit values obtained from matrix sizes between $N=100$
and $N=1000$ (squares) or $N=2400$ (circles). 
The dashed line is the theoretical prediction for the
transition from $\nu  = 1$, at small $\alpha $, to $\nu  = 0$, at large
$\alpha $.
\\[3mm]

\noindent
{\bf Fig. 4} \\
The same as Fig. 3, but for the index $\mu $, derived from the $N$ dependence
of the variance $\delta $ of the inverse participation ratio:
$\delta \equiv (\langle I_2^2\rangle - \langle I_2\rangle^2)/\langle
I_2\rangle^2 \propto 1/N^\mu $. The dashed line corresponds to
the predicted linear crossover from $\mu =1$ at $\alpha < 3/4$ to $\mu =0$ at
$\alpha > 1$.
\\[3mm]

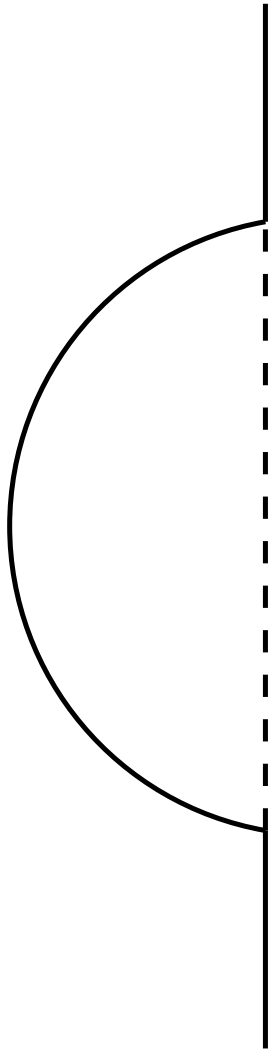
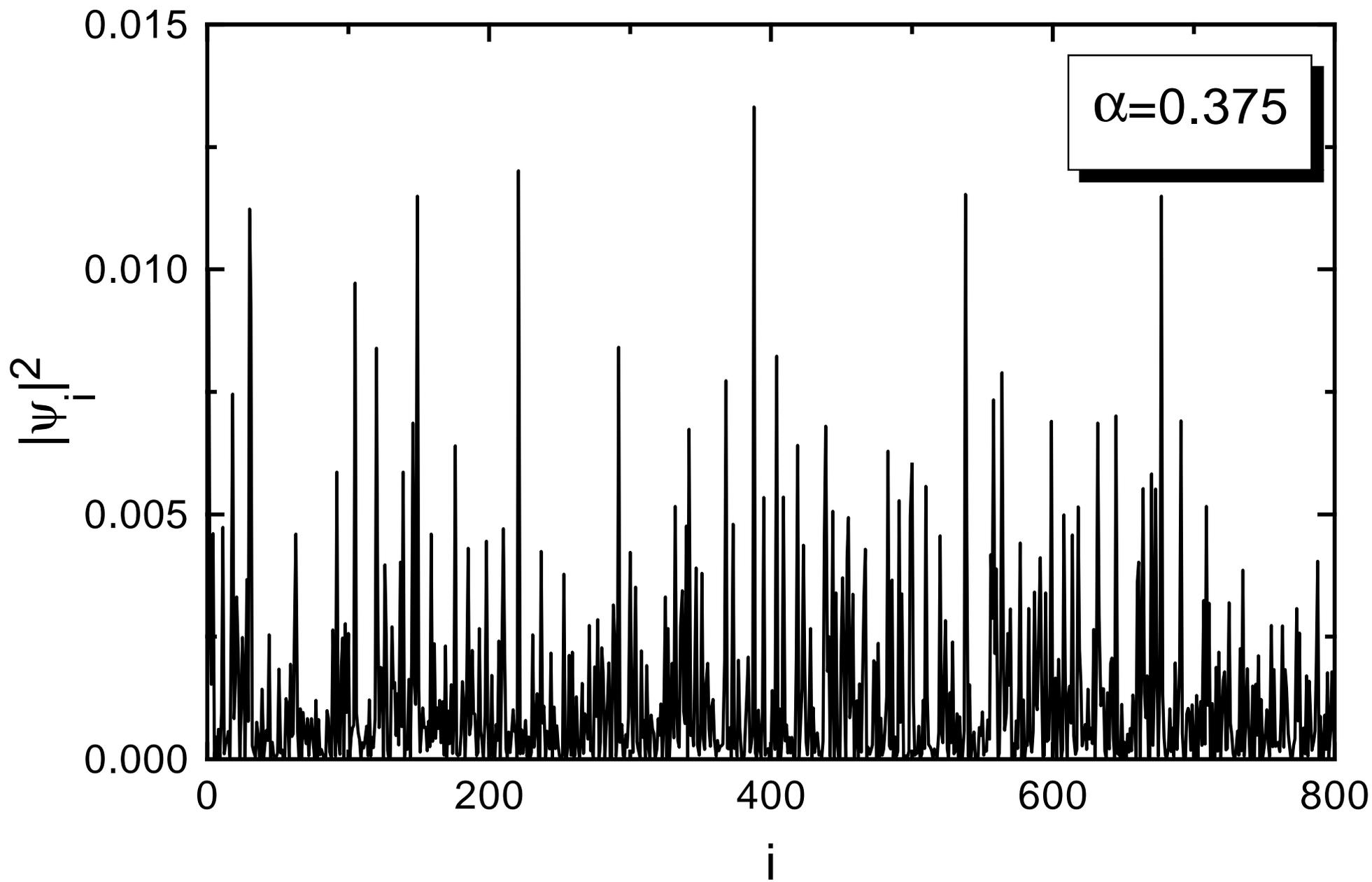
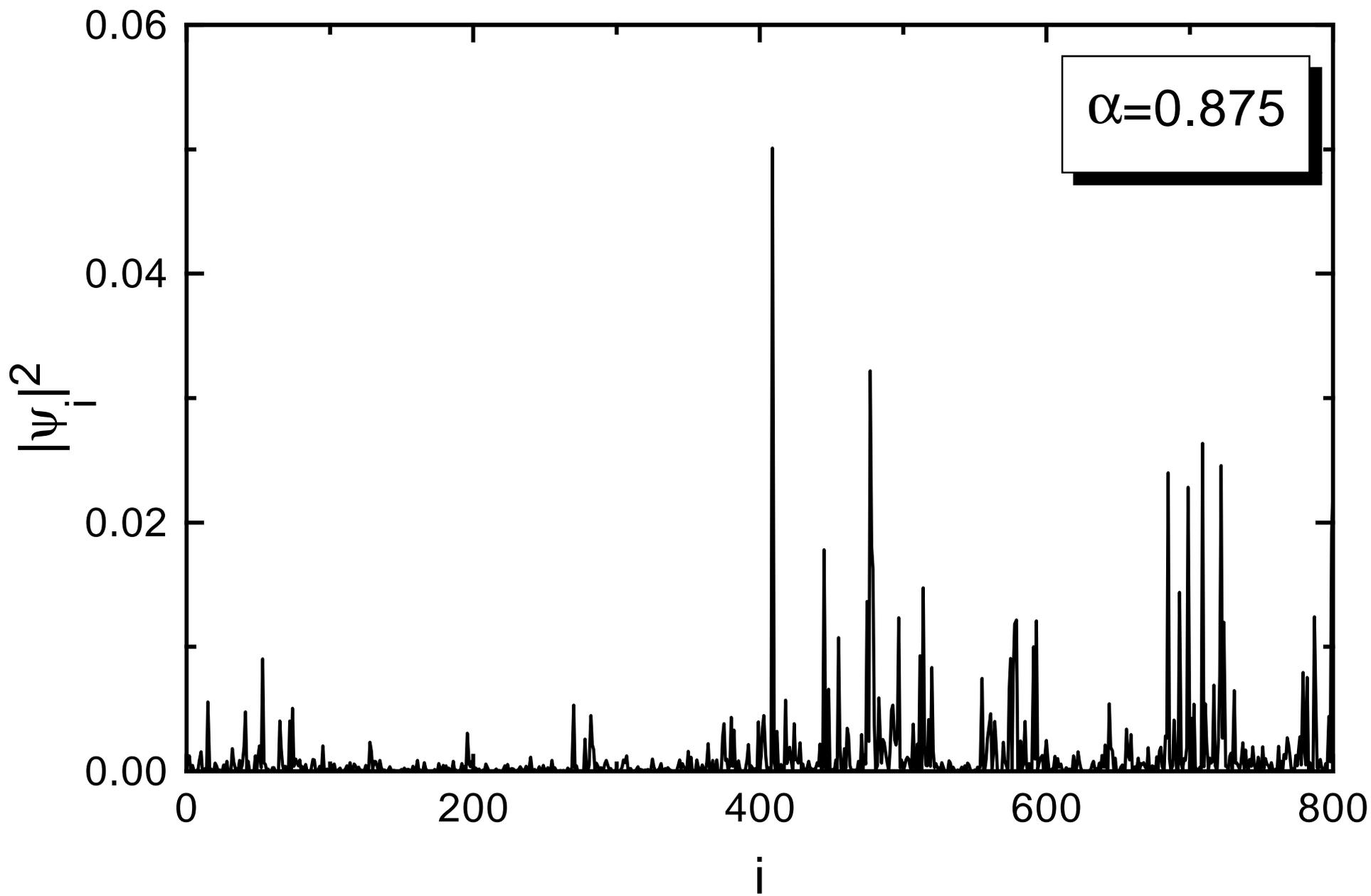
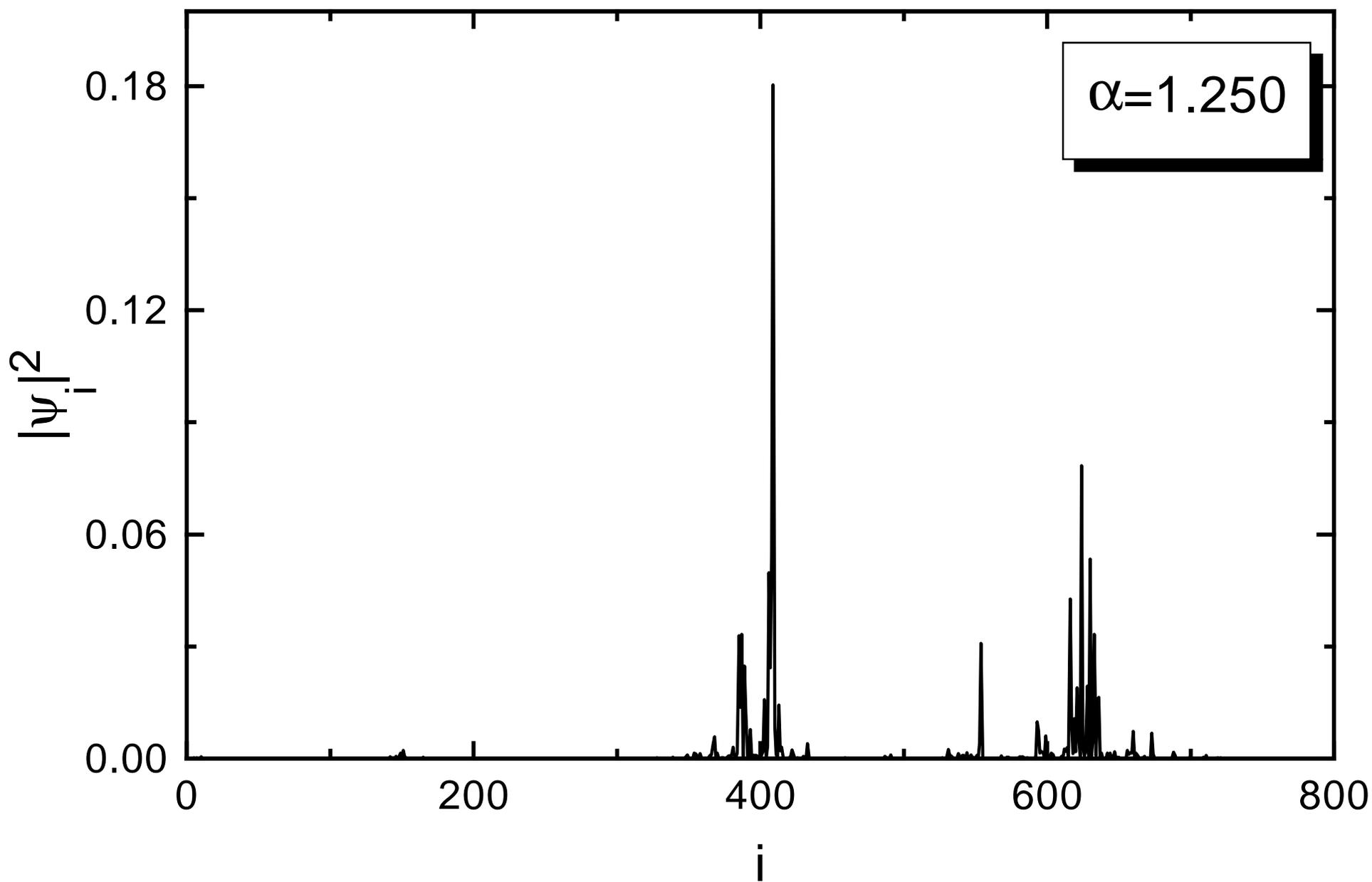
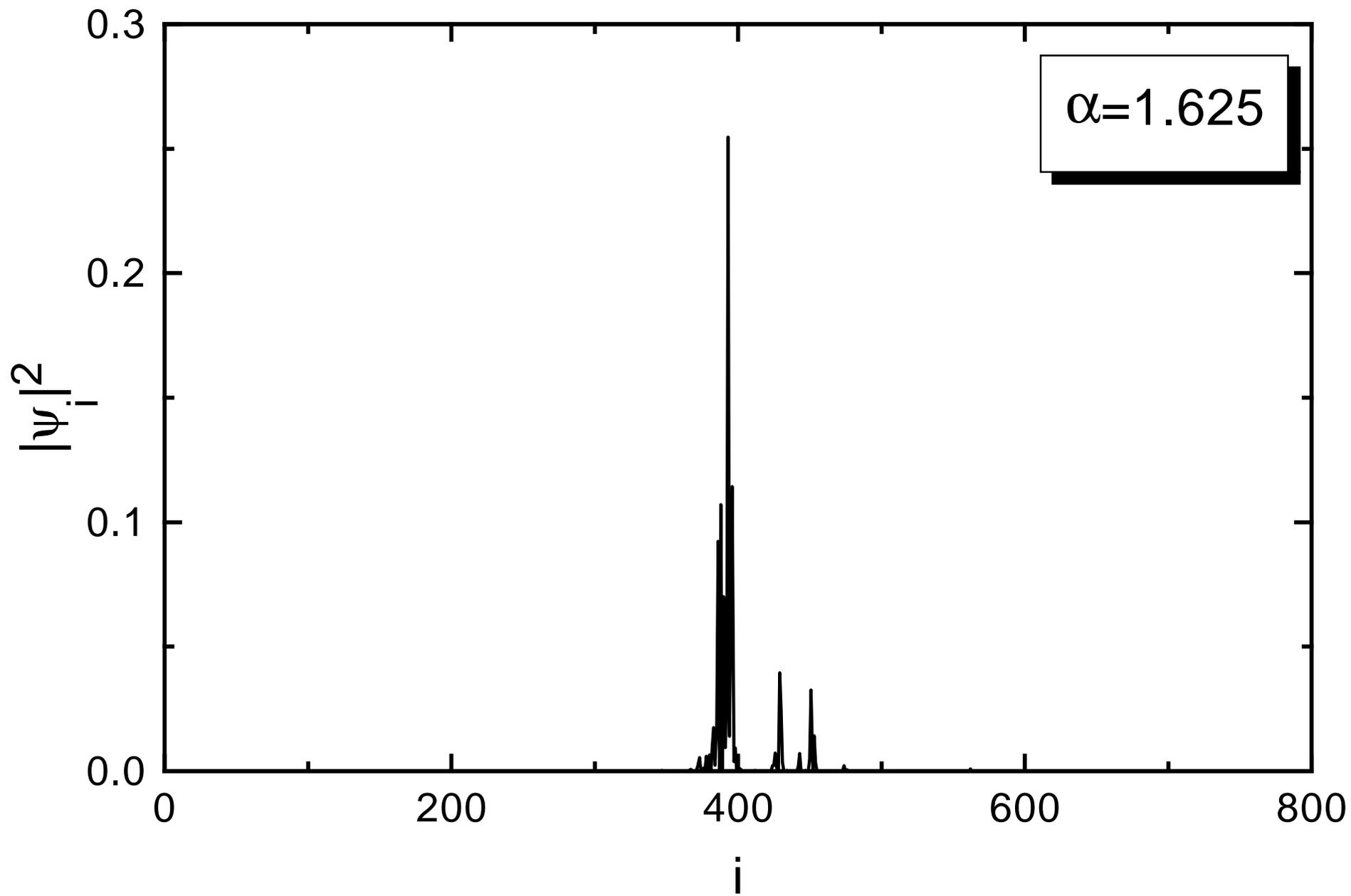
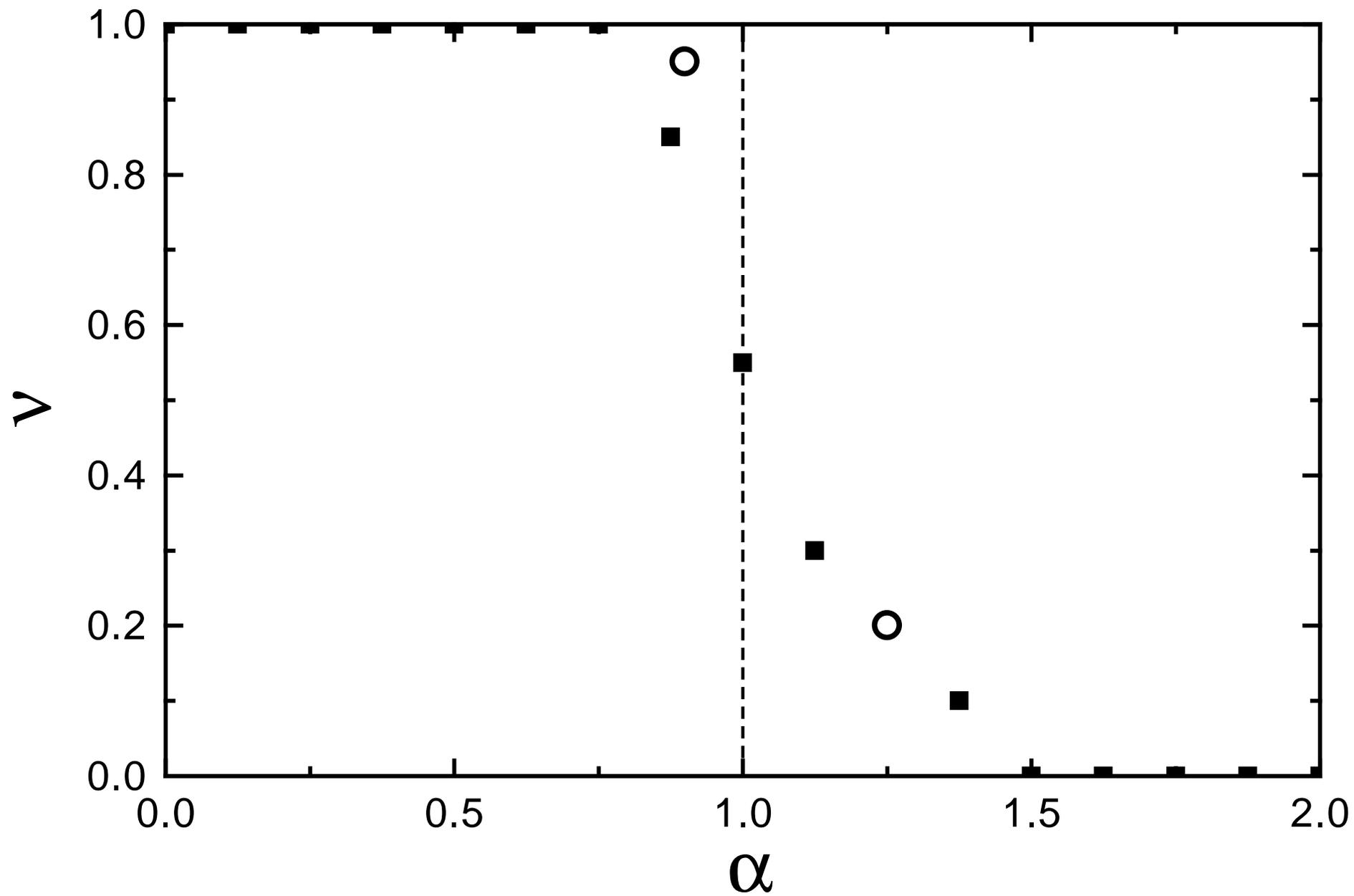
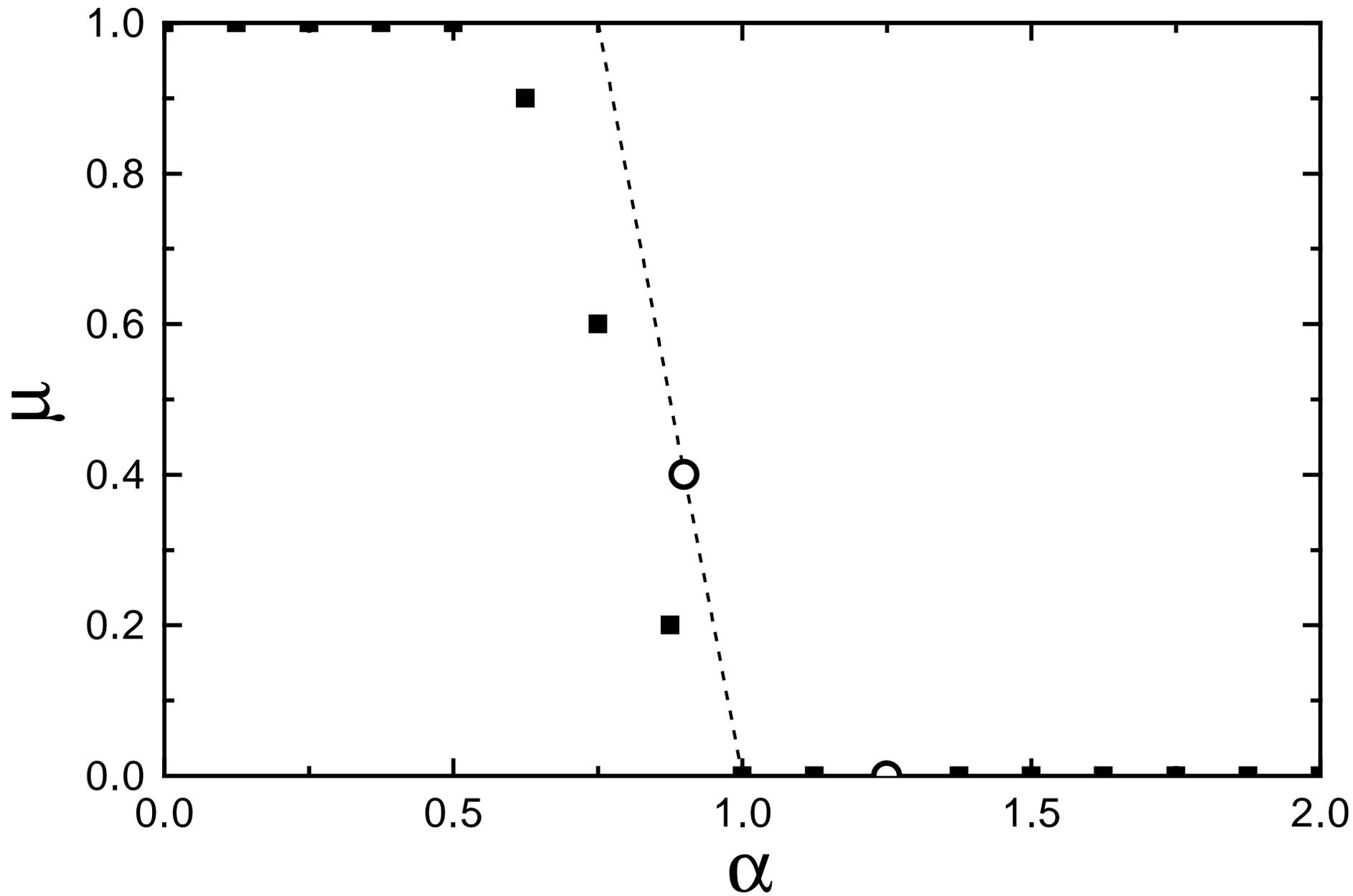
\end{document}